\documentclass[
superscriptaddress,
]{revtex4}

\usepackage[latin9]{inputenc}
\usepackage{diagbox}
\usepackage{soul}
\usepackage{amsmath,amsfonts,mathrsfs,amssymb}
\usepackage{epsfig}
\usepackage{latexsym}
\usepackage{pdfsync}
\usepackage{graphicx}
\usepackage{dcolumn}
\usepackage{bm}
\usepackage{bbm}
\usepackage[T1]{fontenc} 
\usepackage{epsf}
\usepackage{amsthm}
\usepackage{color}
\usepackage{slashed}
\usepackage{float}
\usepackage{esvect}
\usepackage{mathtools}
\usepackage{hyphenat}
\usepackage[pdfencoding=auto]{hyperref}
\newcommand{\RNum}[1]{\uppercase\expandafter{\romannumeral #1\relax}}

\begin{document}

\title{The hadron spectra and pion form factor in dynamical holographic QCD model with anomalous 5D mass of scalar field}
	
\author{Ruixiang Chen}
\email[]{chenruixiang22@mails.ucas.ac.cn}
\affiliation{School of Nuclear Science and Technology, University of Chinese Academy of Sciences, Beijing 100049, China}	

\author{Danning Li}
\email[]{lidanning@jnu.edu.cn}
\affiliation{Department of Physics and Siyuan Laboratory, Jinan University, Guangzhou 510632, P.R. China}
	
\author{Kazem Bitaghsir Fadafan}
\email[]{bitaghsir@shahroodut.ac.ir}
\affiliation{ Faculty of Physics, Shahrood University of Technology, P.O.Box 3619995161 Shahrood, Iran}

\author{Mei Huang}
\email[]{huangmei@ucas.ac.cn}
\affiliation{School of Nuclear Science and Technology, University of Chinese Academy of Sciences, Beijing 100049, China}	
	
\begin{abstract}
		
\noindent
The simplest version of the dynamical holographic QCD model is described by adding the KKSS model action on a dilaton-graviton coupled background, in which the AdS$_5$ metric is deformed by the gluon condensation and further deformed by the chiral condensation. In this framework, both the chiral symmetry breaking and linear confinement can be realized, the light-flavor hadron spectra and the pion form factor were investigated but it was difficult to reconcile the light-flavor hadron spectra and pion form factor. By considering the anomalous 5-dimension mass correction of the scalar field from QCD running coupling, it is found that the light flavor hadron spectra and pion form factor can be described well simultaneously, especially the ground state and lower excitation states of scalar, pseudo scalar and axial vector meson spectra are improved, but the vector meson spectra is not sensitive to the anomalous 5-dimension mass correction of the scalar field.
	
\end{abstract}
	
\maketitle
\section{Introduction}

As the fundamental theory of strong interaction describing more than 99$\%$ of visible matter in the universe, quantum chromodynamics (QCD) is quite successful in the ultraviolet region when the coupling is weak \cite{Gross:1973id, Politzer:1973fx}. However, solving nonperturbative QCD physics in the infrared (IR) regime still remains as a challenge in hadron physics and QCD phase transitions related to the chiral symmetry breaking and color confinement.  In recent decades, the holographic QCD method based on the anti-de Sitter/conformal field theory (AdS/CFT) correspondence or gauge/gravity duality \cite{Maldacena:1997re,Gubser:1998bc,Witten:1998qj} has become an important tool in dealing with nonperturbative QCD problems, and has been widely applied in hadron physics and strongly coupled quark matter.

Based on the conjecture of AdS/CFT correspondence, it is widely believed that there exists a general holography between a D-dimensional quantum field theory (QFT) and a (D + 1)-dimension quantum gravity, with the extra dimension as an emergent energy scale or renormalization group (RG) flow in the QFT \cite{Adams:2012th}. There have been lots of efforts to construct a non-conformal 5-dimensional holographic QCD model from both top-down and bottom-up. For example, the $D_p-D_q$ system from top-down including the $D_3-D_7$ \cite{Erdmenger:2007cm} and the $D_4-D_8$ system or the Witten-Sakai-Sugimoto (WSS) model \cite{Sakai:2004cn,Sakai:2005yt}. In the bottom-up approach, the hard-wall model \cite{Erlich:2005qh} and soft-wall AdS/QCD model or KKSS model \cite{Karch:2006pv} established the 5-dimensional framework for light hadron spectra. In the simplest models, like the KKSS model, the generation of the quark condensate has not been included and the decoupling of the deep IR from the physics of the mesons is dictated by hand using a soft wall at IR. Progresses were made in Refs. \cite{Colangelo:2008us, Ghoroku:2005vt,Gherghetta:2009ac,Sui:2009xe}.

In the bottom-up method, in order to break the conformal invariance, energy scale dependent operators including IR physics are taken into account. For example, in \cite{Cui:2013xva}, a z-dependent mass has been introduced so that IR physics can be improved in the soft-wall AdS/QCD model.  The authors in recent paper \cite{Vega:2021yyj} discussed the effects of beta function on mass and melting temperature for scalar glueballs. The chiral condensate has been studied from a general point of view in  \cite{Cherman:2008eh} . In \cite{Gherghetta:2009ac},  the dilaton profile and the potential in the KKSS model action have been modified by a quartic term to get the chiral symmetry breaking. In \cite{Vega:2010ne}, the scalar mass is dependent on the holographic radial coordinate to study the meson spectra in the model, and the generalized study to consider interactions has been done in \cite{Vega:2011tg}.

The anomalous dimension has also been introduced in bottom-up holographic models to break conformal symmetry. In the Gubser model \cite{Gubser:2008yx,Gubser:2008ny,DeWolfe:2010he,DeWolfe:2011ts}, an anomalous dimension of gluon condensation has been taken into account. The beta function for scalar gluballs has been introduced in \cite{Boschi-Filho:2012ijd}. Scaling Dimensions and the gluon field propagator has been studied in \cite{Powell:2012zj}.
A warp factor related to photon field has been introduced in \cite{MartinContreras:2021yfz} for the electromagnetic pion form factor as well as the pion radius. In \cite{FolcoCapossoli:2022jtr},  an anomalous dimension of an operator for fermions at the boundary has been taken into account to compute the corresponding structure functions. In the D3/D7 system, it has been phenomenologically adjusted to include a running anomalous dimension $\gamma$ for the quark quark condensate \cite{BitaghsirFadafan:2018efw,Belyaev:2019ybr,BitaghsirFadafan:2019ofb}. From the AdS/CFT correspondence, the non zero quark condensate is dual to the scalar field in the AdS background suffering an instability. When the scalar field mass is $m^2=-4$, it passes through the Breitenlohner-Freedman bound \cite{Breitenlohner:1982jf}.

Systematic framework such as the improved holographic QCD (IhQCD) model \cite{Gursoy:2007cb,Gursoy:2007er}, the V-QCD model \cite{Jarvinen:2011qe} and the dynamical holographic QCD model \cite{Chen:2022goa,Li:2013oda} have been developed to incorporate linear confinement and chiral symmetry breaking. The improved holographic QCD (IhQCD) model \cite{Gursoy:2007cb,Gursoy:2007er} and the V-QCD model embedded in the running coupling from UV to IR in the dilation potential.
In the simplest version of the dynamical holographic QCD model \cite{Chen:2022goa,Li:2013oda},  the KKSS model action describing the flavor physics is added on a dilaton-graviton coupled background. Thus the AdS$_5$ metric is automatically deformed by the gluon condensation and further deformed by the chiral condensation. In this framework, both the chiral symmetry breaking and linear confinement can be realized, therefore, in the produced light-flavor hadron spectra, one can read the information of chiral symmetry breaking: the $140 {\rm MeV}$ pseudo Nambu-Goldstone boson,i.e., the pion, and the splitting between the chiral partners, as well as the linear confinement information through the linear Regge behavior of higher excitation states. It was difficult to reconcile the light-flavor hadron spectra and pion form factor in \cite{Li:2013oda}. To improve this part, in this work, we consider the anomalous 5-dimension mass correction of the scalar field from QCD running coupling. In this paper we present an extension of model so that it includes the dynamics of gauges. We input it through an assumption for the running of the anomalous dimension of the quark antiquark, $\gamma$. We take the running of $\gamma$ from the perturbative two loop result for $SU(3)$ gauge theory with two flavors $N_f=2$.

It is found that with the anomalous 5-dimension mass correction of the scalar field, the light flavor hadron spectra and pion form factor can be described well simultaneously, especially the ground state and lower excitation states of scalar, pseudo scalar and axial vector meson spectra are improved, but the vector meson spectra is not sensitive to the anomalous 5-dimension mass correction of the scalar field. The paper is organized as following. After introduction, in Sec.\ref{sec-model}, we will briefly introduce the dynamical holographic QCD model, then introduce the anomalous dimension of scalar field in section \ref{sec-anom-dimension} and show numerical results on hadron spectra and pion form factor
in Sec.\ref{sec-results}. At last, we will give summary and discussion.

\section{Dynamical holographic QCD model}
\label{sec-model}
Following Refs.~\cite{Li:2013oda,Li:2012ay}, one can couple the Einstein-Dilaton system and the KKSS action, which are assumed to describe the gluonic dynamics and the flavor dynamics respectively. Then, we have the following system
\begin{equation}
S = S_{G}+ \frac{N_f}{N_c} S_{f},
\end{equation}
with the gluon background part $S_{G}$ and flavor part $S_{f}$ of the following form
\begin{eqnarray}
&&\!\!\!\!\!\!\!\!\!\!\!\!\!\!\!\! S_{G} = \frac{1}{16\pi G_5}\int d^5x\sqrt{g_s}e^{-2\Phi}\big(R+4\partial_M\Phi\partial^M\Phi-V_G(\Phi)\big), \\
 &&\!\!\!\!\!\!\!\!\!\!\!\!\!\!\!\! S_{f}=-\int d^5x e^{-\Phi(z)} \sqrt{g_s}Tr\Big(|DX|^2+V_X(|X|,\Phi)
 +\frac{1}{4g_5^2}(F_L^2+F_R^2)\Big) \label{action-KKSS}.
\end{eqnarray}
Here we have taken $S_f=S_{KKSS}$ the simplest version for the flavor part. In Eq.(\ref{action-KKSS}), $\Phi$ is the dilaton field,  $X\equiv X^{\alpha \beta}$ is a matrix-valued scalar field with $\alpha,\beta$ the flavor indexes, $g_s \equiv  \text{det}(g^s_{\mu\nu})$ is the determinant of the metric in string frame, and $V_X$ is the scalar potential which includes the interaction between $X$ and $\Phi$ (strictly speaking, it is an extension of the original KKSS model).  $F_L$ and $F_R$ represent the gauge strength of the left-handed guage potential $L^{M}= L^{Ma} t^a$  and right-handed gauge potential $R^{M}= R^{Ma} t^a$
\begin{eqnarray}
F_{L}^{MN}&=&\partial^{M}{L^{N}}-\partial^{N}{L^{M}}-i[L^{M},L^{N}],\nonumber\\
F_{R}^{MN}&=&\partial^{M}{R^{N}}-\partial^{N}{R^{M}}-i[R^{M},R^{N}],
\end{eqnarray}
with $t^a$ the generators of the group considered, for which we have Tr$[t^{a}t^{b}]=\delta^{ab}/2$. One can transform the lef-handed gauge field $L$ and the right-handed gauge field $R$ into the vector ($V$) and axial-vector ($A$) fields with
$L^{M}=V^{M}+A^{M}$ and $ R^{M}=V^{M}-A^{M}$. $DX$ represents the covariant derivative, which takes the form
\begin{equation}
D^M X=\partial^M X-i L^M X+iX R^M.
\end{equation}
One of the main tasks to build the holographic model is to solve the gravity background which is dual to the vacuum. In this work, we consider all the current quarks are degenerate. Thus, we take a simple ansatz for the background part of $X$, i.e. $X=\frac{\chi}{2} I$, with $I$ the identity matrix. Since the vacuum does not carry other no-zero quantum numbers, we assume the gauge fields to be zero in the vacuum and consider them as perturbations above the vacuum. Therefore, in the background part, only the metric, the dilaton field, and the diagonal part of the scalar field $\chi$ appear. Inserting the above ansatz into the action, we get an effective expression in terms of $\Phi$ and $\chi$ as
\begin{eqnarray}
S_{vac} \!\!\! &=& \!\!\! \frac{1}{16 \pi G_5} \int d^5 x \sqrt{g_s} \big\{ e^{-2\Phi}[R
+4\partial_M\Phi \partial^M \Phi - V_G(\Phi)] \nonumber\\
&-& \lambda e^{-\Phi} (\frac{1}{2} \partial_M\chi \partial ^M \chi
+ V_C(\chi,\Phi))\big\}.
\end{eqnarray}
Here, we have redefined the coupling between the two sectors as $\lambda\equiv \frac{16\pi G_5 N_f}{ N_c}$ (for simplicity, we will take the AdS radius $L=1$).
Considering the Lorenzian symmetry of the vacuum, one could take the ansatz of the metric as
\begin{eqnarray}\label{metric-ansatz}
g^s_{MN}=b_s^2(z)(dz^2+\eta_{\mu\nu}dx^\mu dx^\nu), ~ ~ b_s(z)\equiv e^{A_s(z)}.
\end{eqnarray}
Here, $z$ is the holographic dimension and  $z=0$ is the 4D boundary. $\eta_{\mu\nu}$ is the metric of Minkowski space with $\eta_{00}=-1$.
Then, it is not difficult to derive the equation of motion from Einstein equations and the field equations, and it reads
\begin{eqnarray}
 -A_s^{''}+A_s^{'2}+\frac{2}{3}\Phi^{''}-\frac{4}{3}A_s^{'}\Phi^{'}
 -\frac{\lambda}{6}e^{\Phi}\chi^{'2}&=&0, \label{Eq-As-Phi} \\
 \Phi^{''}+(3A_s^{'}-2\Phi^{'})\Phi^{'}-\frac{3\lambda}{16}e^{\Phi}\chi^{'2}
 -\frac{3}{8}e^{2A_s-\frac{4}{3}\Phi}\partial_{\Phi}\left(V_G(\Phi)
 +\lambda e^{\frac{7}{3}\Phi}V_C(\chi,\Phi)\right)&=&0, \label{Eq-VG}\\
 \chi^{''}+(3A_s^{'}-\Phi^{'})\chi^{'}-e^{2A_s}V_{C,\chi}(\chi,\Phi)&=&0. \label{Eq-Vc}
\end{eqnarray}

According to the analysis in Refs.\cite{Li:2013oda,Li:2012ay}, in order to incoorperate the linear spectrum, the linear potential, and the chiral symmetry breaking, one could take the following field configurations,
\begin{eqnarray}\label{phiz}
&&\Phi(z)=\mu_G^2z^2,\label{chiz}\\
&&\chi^{'}(z)=\sqrt{8/\lambda}\mu_G e^{-\Phi/2}(1+c_1 e^{- \Phi}+c_2
e^{-2\Phi}),
\end{eqnarray}
with $\mu_G$ a model parameter related to the Regge slope of the spectrum. The quadratic form of the dilaton field is taken from the original KKSS model. It turns out to be responsible for the linear spectrum of light mesons. The specific form of $\chi^\prime$ is introduced phenomenologically, which leads $A_s^\prime\rightarrow 0, A_s\rightarrow Const$ when $z$ approaches infinity, and is responsible for the linear part of the quark-antiquark potential (the linear confinement from another point of view). Furthermore, in order to describe the chiral condensate, the asymptotic behavior of $\chi$ should take the following form,
\begin{eqnarray}
\chi(z) \stackrel{z \rightarrow 0}{\longrightarrow} m_q \zeta z+\frac{\sigma}{\zeta} z^3.
\label{chi-UV}
\end{eqnarray}
Thus, we have the two coefficients to be
$c_1=-2+\frac{5\sqrt{2\lambda}m_q\zeta}{8\mu_G}+\frac{3\sqrt{2\lambda}\sigma}{4\zeta
\mu_G^3},c_2=1-\frac{3\sqrt{2\lambda}m_q\zeta}{8\mu_G}-\frac{3\sqrt{2\lambda}\sigma}{4\zeta
\mu_G^3}$. Here, $m_q$ and $\sigma$ are the current quark mass and chiral condensate respectively. $\zeta$ is a normalization constant, which is introduced by matching the 5D results of the two-point function of the scalar operator $
\bar{q}q$ to the 4D perturbative calculations \cite{Cherman:2008eh}.

Having all these considrations, one could take proper values of the parameters $\mu_G, m_q, \sigma,\lambda$ and solve the metric, and the dilaton and scalar potentials $V_G, V_C$ from the equations of motion. Generally, they are functions of $z$, like $\Phi, \chi$. After solving all the functions, one might extract the dilaton and the scalar potential $V_G, V_{c,\chi}$. To do so, we assume the interaction between $\Phi$ and $\chi$ takes a simple form $V_{c,\chi}=e^{f(\Phi)}V_{c,\chi}(\chi)$ and simply take $f(\Phi)=-\Phi/2+\ln(1+\Phi)/2$. Then one can easy extract the asymptotic behavior of $V_c(\chi)$ as
\begin{eqnarray}
V_{c}(\chi)=\frac{1}{2}M^2\chi^2+o(\chi^2),
\end{eqnarray}
with the 5D mass $M^2=-3$, which satisfies the operator dimension $\Delta$ and 5D mass relation $M^2=\Delta(\Delta-4)$ for scalar operator $\bar{q}q$ with $\Delta=3$.


\section{Anomalous dimension corrections }
\label{sec-anom-dimension}

The above form could be considered as defined at ultra-UV scale, where the dimension of $\bar{q}q$ is $\Delta=3$. However, according to the perturbative calculation in 4D QCD, the dimension of the scalar operator $\bar{q}q$ would obtain corrections, i.e. the anomalous dimension $\gamma$, when considering quantum fluctuations.  Generally, the anomalous dimension would depend on the energy scale. Considering the duality is valid at strong coupling other than at ultra-UV, one might expect an anomalous dimension correction to the operator dimension.

With a nonzero anomalous dimension $\gamma$, the dimension of the operator $\bar{q}q$ becomes $\Delta-\gamma$. From the relation between the 5D mass and the operator dimension, the mass of the dual 5D scalar field would be
\begin{equation}\label{M-Delta}
    \tilde{M}^2=M^2+\Delta M^2=(\Delta-\gamma)(\Delta-\gamma-4).
\end{equation}
Correspondingly, the asymptotic expansion of $\chi$, Eq.\eqref{chi-UV} becomes
\begin{eqnarray}
\chi(z) \stackrel{z \rightarrow 0}{\longrightarrow} m_q \zeta z^{1+\gamma}+\frac{\sigma}{\zeta} z^{3-\gamma}.
\label{chi-UV-constant}
\end{eqnarray}

Phenomenologically, one can take a constant value of $\gamma$ as in Ref.\cite{Gubser:2008yx}, work out $\Delta M^2$ and solve the corrections to the background metric and fields. To be more careful, the anomalous dimension is generally a function of the energy scale. In the 5D description, since the holographic direction could be mapped to the energy scale, one might transfer the energy scale dependence of the anomalous dimension to the 5th dimension evoloution of the anomolous dimension. Thus, $\gamma$ should be a function of the holographic dimension, i.e. as a function $\gamma(z)$ other than a constant. Then, the asymptotic expansion of $\chi(z)$ would become
\begin{eqnarray}
\chi(z) \stackrel{z \rightarrow 0}{\longrightarrow} m_q \zeta z^{1+\gamma(z)}+\frac{\sigma}{\zeta} z^{3-\gamma(z)}.
\label{chi-UV-gamma-z}
\end{eqnarray}
The 5D mass then effectively becomes
\begin{equation}\label{M-Delta}
    \tilde{M}^2=[\Delta-\gamma(z)][\Delta-\gamma(z)-4].
\end{equation}
Then, if the 5D mass in the potential $V_{C,\chi}(\chi,\Phi)$ is modified as the above form, the corrections from the anomalous dimension could be solved. However, since $V_{C,\chi}(\chi,\Phi)$ is numerically obtained in the previous works Refs.\cite{Li:2013oda,Li:2012ay}, it is not directly to add the correction in this way. Instead, we will introduce the corrections in an effective way. Comparing Eq.\eqref{chi-UV-gamma-z}  with  Eq.\eqref{chi-UV}, one might expect a minor corrections to the previous model, by replacing the constant $c_1, c_2$ as functions of $z$ . Therefore, we would have
\begin{eqnarray}
&&\Phi(z)=\mu_G^2z^2,\label{phiz-1}\\
&&\chi^{'}(z)=\sqrt{8/\lambda}\mu_G e^{-\Phi/2}(1+c_1(z) e^{- \Phi}+c_2(z)
e^{-2\Phi})\label{chiz-1}.
\end{eqnarray}
The functions $c_1(z)$ and $c_2(z)$ could be obtained by comparing the UV asymptotic expansion of Eq.\eqref{chiz-1} to Eq.\eqref{chi-UV-gamma-z}, and they have the following form
\begin{eqnarray}
c_1&=&-2+\frac{5\sqrt{2\lambda}m_q e^2 z^{\gamma(z)}\zeta}{8\mu_G}+\frac{3\sqrt{2\lambda}\sigma  z^{-\gamma(z)}}{4e^2\zeta
\mu_G^3},\\
c_2&=&1-\frac{3\sqrt{2\lambda}m_qe^2 z^{\gamma(z)}\zeta}{8\mu_G}-\frac{3\sqrt{2\lambda}\sigma z^{-\gamma(z)}}{4e^2\zeta
\mu_G^3}
\end{eqnarray}
Here, the factor $e^2$ is to guarantee  that at ultra UV the dimension recovers 3. In this way, the asymptotic behavior of $\chi$ would become Eq.\eqref{chiz-1}, and the anomalous dimension is effectively introduced.

To solve the dependence of $z$ of $\gamma(z)$, one has to consider the constraints from the 4D perturbative calculations. The running of the gauge coupling in QCD theory is controlled by the $\beta$ function as
\begin{eqnarray}\label{betaeq}
	\mu { d \alpha \over d \mu} = \beta(\alpha) .
\end{eqnarray}

From the 4D one loop results, the anomalous dimension of the operator $\bar{q}q$ takes the form of
\begin{equation}
	 \gamma = {3 (N_c^2-1) \over 4 N_c \pi} \alpha\,.  \end{equation}

We assume that the inverse of the renormalization scale $\mu^{-1}$ corresponds to the $z$ coordinate and Eq.\eqref{betaeq} becomes

\begin{equation}
-z \frac{d \alpha}{dz}=\beta(\alpha).
 \end{equation}

In this work, we consider the one loop perturbative calculation only, so we have
\begin{eqnarray}
\beta(\alpha)= - b_0 \alpha^2 ,
\end{eqnarray}
with
\begin{equation} b_0 = {1 \over 6 \pi} (11 N_c - 2N_f). \end{equation}
The perturbative results are valid at high energy scales only. To extend it to IR region, we impose a simple extrapolation following Ref.\cite{Alanen:2009na,FolcoCapossoli:2016uns,Rodrigues:2016kez}, which reads
\begin{eqnarray}
\beta(\alpha)= - b_0 \alpha^2(1-\alpha/\alpha_0)^2 .
\end{eqnarray}
Here, $\alpha_0$ is the IR fix point and would be considered as a model parameter in this work. Taken the above form, one can solve the beta function and the $z$ dependence of $\gamma$, and then solve the equations of motion Eqs.(\ref{Eq-As-Phi},\ref{Eq-VG},\ref{Eq-Vc}).

\begin{figure}[ht!]
    \centering
    \includegraphics[width=0.5\linewidth]{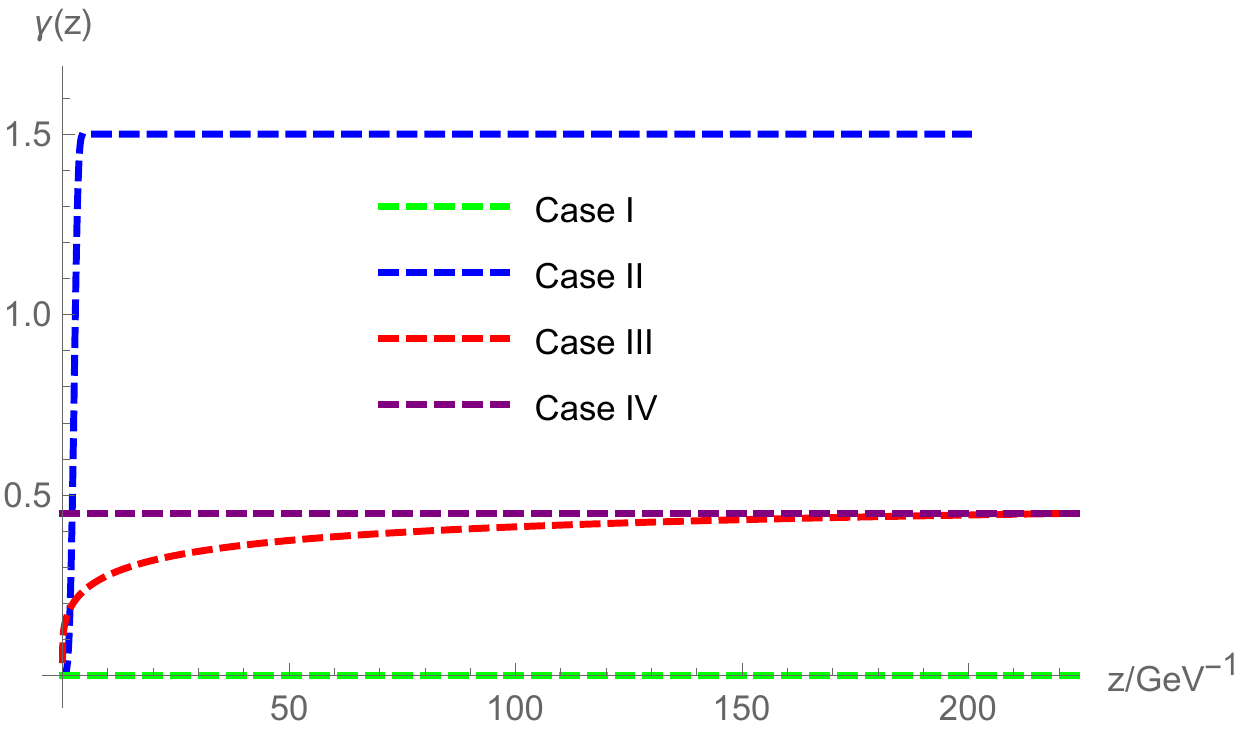}
    \caption{\label{fig-gamma}The anomalous dimension as a function of the holographic dimension $\gamma(z)$. }
\end{figure}

Given that the non-perturbative dominant physics for hadrons, one might expect the UV perturbative part contributes less to the final results, thus we also consider two simple models of the anomalous dimensions for comparisons: a). we take $\gamma(z)=-\frac{3}{2}(e^{-\Phi^2/2}-1)$ and replace $m_q,\sigma$ with $m_q z^{\gamma(z)}[\frac{1}{2}(1+\gamma(z))+\ln(z)\gamma^\prime(z)], \sigma z^{-\gamma(z)}[\frac{1}{2}(3-\gamma(z))-\ln(z)\gamma^\prime(z)]$; b). we also take a constant $\gamma$.

\begin{figure}[ht!]
    \centering
    \includegraphics[width=0.4\linewidth]{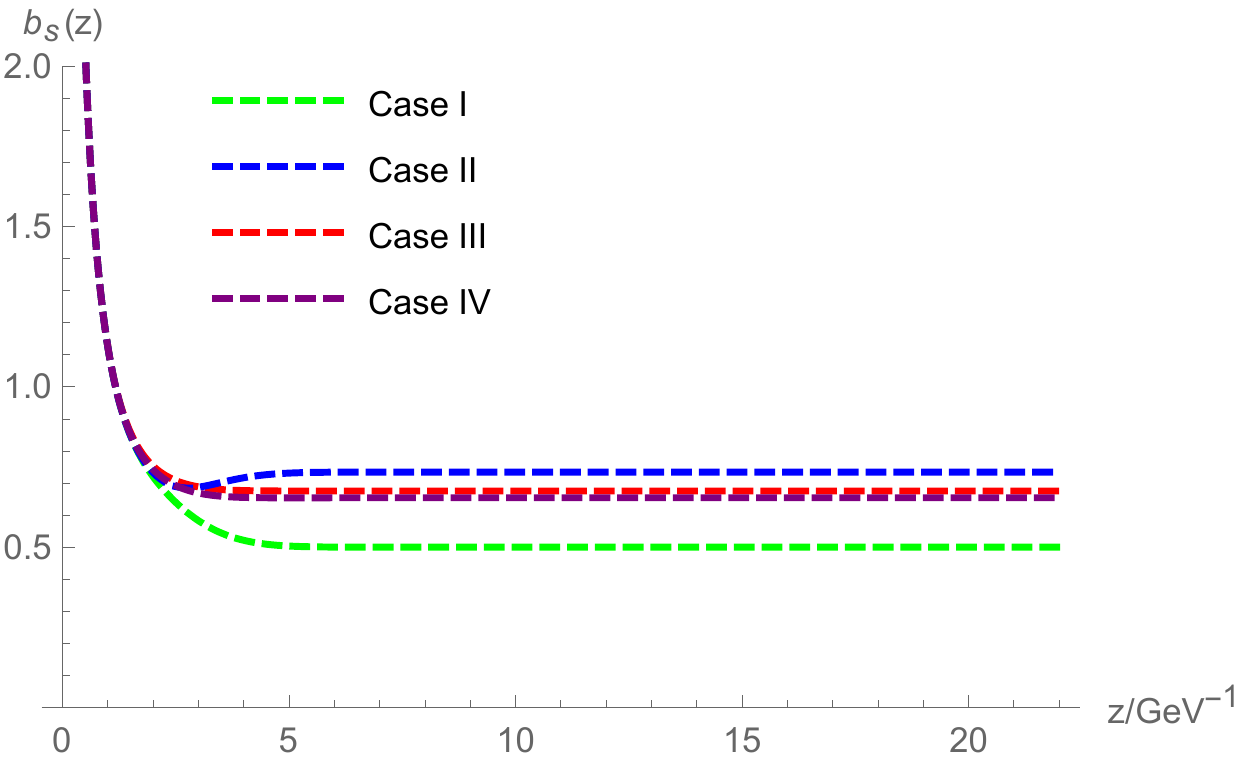}
    \hspace*{1cm}
    \includegraphics[width=0.4\linewidth]{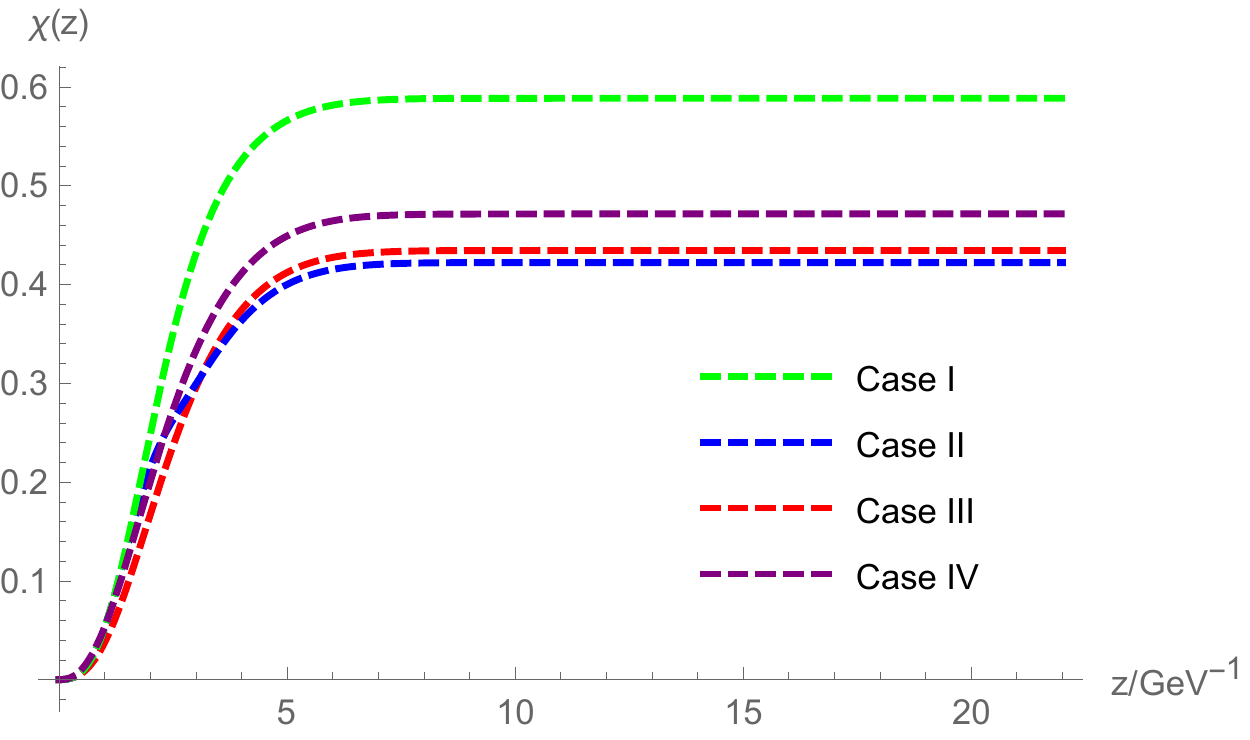}
     \vskip -0.05cm \hskip 0 cm
\textbf{( a ) } \hskip 8 cm \textbf{( b )}
    \caption{\label{fig-bs-chi}The warp factor $b_s(z)$ (a) and the scalar field $\chi(z)$ (b). }
\end{figure}

Finally, to solve the models, we need to fix the parameters. Firstly, in this work, we will focus on the case $N_c=3, N_f=2$. The value of the parameter $\mu_G$ describes the Regge slope and it could be fixed as $\mu_G=0.43\rm{GeV}$. Follwoing Ref.\cite{Li:2013oda}, we will take $G_5=0.75, m_q=5\rm{MeV}, \sigma=(240\rm{MeV})^3$ (the model IB in Refs.\cite{Li:2013oda,Li:2012ay}). The coupling at IR fixed point $\alpha_0$ is taken to be $1$. With these parameters, taking the initial condition $\alpha(z=1/M_Z)=0.1184$, one can solve the dynamic $\gamma(z)$ from the beta function as shown in the red dashed line in Fig.\ref{fig-gamma}. Inserting the solution of $\gamma(z)$ into the equation of motion, one could solve the warp factor of the metric $A_s$ and the scalar field. The results are given in the red dashed lines in Fig.\ref{fig-bs-chi}.

As mentioned above, besides the dynamic $\gamma(z)$ from the beta function, we also considered the other three cases. Here, we call the case without anomalous dimension corrections as `Case I'(the same as the Model IB in Ref.\cite{Li:2013oda}), the case with modelling $\gamma(z)=-\frac{3}{2}(e^{-\Phi^2/2}-1)$ as `Case II', the case with the dynamical evolution of $\gamma(z)$ from beta function as `Case III', and the case with a constant $\gamma=0.45$ (around the dynamical value at large $z$) as `Case IV'. The behaviors of $\gamma(z)$ in different model are given in Fig.\ref{fig-gamma}. The solutions of $b_s$ and $\chi$ are obtained by numerically solving the equation of motion, and are shown in Fig.\ref{fig-bs-chi}. From the figure, one could easily see that $b_s$ in all the cases behaves as $b_s\simeq 1/z$, guaranteeing the asymptotic AdS condition. At large $z$, i.e. the IR region, $b_s$ tends to different constants in different models ($0.5,0.73,0.68, 0.65$ for Case I, II, III, IV respectively), which is responsible for the linear part in the quank potential. At UV, the solutions of $\chi(z)$ are constrained by the UV expansion. At IR, it also tends to constants for the different models. Roughly, one can numerically check the values of $b_s \chi$ at large z are around $0.3$ for the different models.

Up to now, we have modelled the anomalous dimension in the dynamical holographic QCD model, and we have solve the background fields describing the vacuum properties. Then, we will continue to study the excitations on the vacuum.

\section{Meson spectra, form factors, decay constants}
\label{formulae}
After solving the background metric and fields. One can study the properties of the mesons as  perturbations on the vacuum.  For the scalar channel, we would take $X=(\frac{\chi}{2}+s)e^{i 2\pi^a t^a}$, with $s, \pi$ the scalar and pseudo-scalar perturbation respectively, considering $s,\pi$ as perturbations. Since the vaccum does not have any background field in the vector channel, we would simply take the vector perturbation as
$v_\mu$ and axial vector perturbation as $a_\mu$.  Calculating the two-point current-current correlation function of the corresponding field, one can obtain the masses of the mesons from their poles. Equivalently, with an example shown in Appendix A, it could also be obtained by solving the following Schrodinger-like equations for $s,\pi,v_\mu,a_\mu$
\begin{eqnarray}\label{scalar-sn}
& & -s_n^{''}+V_s(z)s_n=m_n^2s_n,  \\
& & -\pi_n''+V_{\pi,\varphi}\pi_n=m_n^2(\pi_n-e^{A_s}\chi\varphi_n),  \\
& &  -\varphi_n''+ V_{\varphi} \varphi_n=g_5^2 e^{A_s}\chi(\pi_n-e^{A_s}\chi\varphi_n), \\
& & -v_n^{''}+V_v(z)v_n=m_{n,v}^2v_n, \label{vector-n} \\
& & -a_n^{''}+V_a a_n = m_n^2 a_n,
\end{eqnarray}
with the potentials
\begin{eqnarray}
&&V_s=\frac{3A_s^{''}-\Phi^{''}}{2}+\frac{(3A_s^{'}-\Phi^{'})^2}{4}+e^{2A_s}V_{C,\chi\chi},
\label{s-vz}\\
&& V_{\pi,\varphi}=\frac{3A_s^{''}-\Phi^{''}+2\chi^{''}/\chi-2\chi^{'2}/\chi^2}{2} \label{ps-vz} +\frac{(3A_s^{'}-\Phi^{'}+2\chi^{'}/\chi)^2}{4},  \\
&& V_{\varphi} = \frac{A_s^{''}-\Phi^{''}}{2}+\frac{(A_s^{'}-\Phi^{'})^2}{4},  \\
&& V_v=\frac{A_s^{''}-\Phi^{''}}{2}+\frac{(A_s^{'}-\Phi^{'})^2}{4} ~, \label{v-vz} \\
&& V_a = \frac{A_s^{'}-\Phi^{'}}{2}+\frac{(A_s^{'}-\Phi^{'})^2}{4}+g_5^2 e^{2A_s}\chi^{2}
\label{a-vz}.
\end{eqnarray}
Here, the longitudinal part $a^{l}_\mu$ of the axial vector perturbations would couple with the pseudo-scalar part, and we have imposed $a^{l}_\mu=\partial_\mu \varphi$. Once one gets the background solution, the masses of the mesons could be obtained by numerically solving the above equations.

It is also easy to extract the decay constants $f_\pi, F_{\rho_n},F_{a_1,n}$ following Refs.\cite{Li:2013oda,Li:2012ay}, and they take the following form
\begin{eqnarray}
f_\pi^2&&=-\frac{N_f}{g_5^2N_c}e^{A_s-\Phi}\partial_z A(0,z)|_{z\rightarrow0}, \\
F_{\rho_n}^2&&=\frac{N_f}{g_5^2N_c}(e^{A_s-\Phi}\partial_z V_n(z)|_{z\rightarrow0})^2, \\
F_{a_1,n}^2&&=\frac{N_f}{g_5^2N_c}(e^{A_s-\Phi}\partial_z A_n(z)|_{z\rightarrow0})^2.
\end{eqnarray}
Here $A(0,z),V_n(z), A_n(z)$ satisfy
\begin{eqnarray}
 (-e^{-(A_s-\Phi)}\partial_z(e^{A_s-\Phi}\partial_z)+g_5^2 e^{2A_s}\chi^2)A(0,z)=0,\\
(-e^{-(A_s-\Phi)}\partial_z(e^{A_s-\Phi}\partial_z)-m_{\rho,n}^2)V_n(z)=0,\\
(-e^{-(A_s-\Phi)}\partial_z(e^{A_s-\Phi}\partial_z)+g_5^2 e^{2 A_s}\chi^2-m_{a_1,n}^2)A_n(z)=0,
\end{eqnarray}
with the boundary condition$A(0,0)=1,\partial_z A(0,\infty)=0$, $V_n(0)=0,\partial_zV_n(\infty)=0$,$A_n(0)=0,\partial_zA_n(\infty)=0$ and normalized as  $\int dz e^{A_s-\Phi} V_m V_n=\int dz e^{A_s-\Phi} A_m A_n=\delta_{mn}.$

The pion form factor can also be derived as Refs.\cite{Li:2013oda,Li:2012ay}
\begin{eqnarray}
  f_\pi^2 F_\pi(Q^2)=\frac{N_f}{g_5^2N_c}\int dz e^{A_s-\Phi} V(q^2,z) \big\{ (\partial_z\varphi)^2+g_5^2\chi^2 e^{2A_s} (\pi-\varphi)^2\big\},
\end{eqnarray}
where$Q^2=-q^2$, and $V(q^2,z),\pi(z),\varphi(z)$ satisfy
\begin{eqnarray}
(-e^{-(A_s-\Phi)}\partial_z(e^{A_s-\Phi}\partial_z
 )+q^2)V(q^2,z)=0,\\
  -e^{-(3A_s-\phi)}\partial_z(e^{3A_s-\phi}\chi^2
 \partial_z)\pi-m_{\pi,n}^2\chi^2(\pi-\varphi)=0, \\
 -e^{-(A_s-\phi)}\partial_z(e^{A_s-\phi}
 \partial_z)\varphi-g_5^2\chi^2e^{2A_s}(\pi-\varphi)=0,
\end{eqnarray}
with the boundary condition $V(q^2,0)=1,\partial_zV(q^2,\infty)=0, \pi(0)=0,\partial_z\pi(\infty)=0,\varphi(0)=0,\varphi(\infty)=0$ and normalized as
\begin{eqnarray}
\frac{N_f}{g_5^2 N_c f_\pi^2}\int dz e^{A_s-\Phi} \big\{ (\partial_z\varphi)^2+g_5^2\chi^2 e^{2A_s} (\pi-\varphi)^2\big\}=1.
\end{eqnarray}

One can also derive the effective coupling between $\rho$ and $\pi$ as
\begin{eqnarray}
g_{n\pi\pi}=g_5 \frac{\int dz e^{A_s-\Phi} V_n \big\{ (\partial_z\varphi)^2+g_5^2\chi^2 e^{2A_s} (\pi-\varphi)^2\big\}}{\int dz e^{A_s-\Phi} \big\{ (\partial_z\varphi)^2+g_5^2\chi^2 e^{2A_s} (\pi-\varphi)^2\big\}},
\end{eqnarray}
with $V, \pi, \varphi$ the wave functions of the corresponding states.

Once one knows the dual background solution, the abvoe equations could be solved and these quantites could be extracted based on the above formulae. We will discuss the results in Sec.\ref{sec-results}.


\section{Numerical results}
\label{sec-results}
After solving the background fields, we can numerically obtain the masses of the mesons, the decay constants and the form factors. As mentioned in Sec.\ref{sec-anom-dimension}, we will consider the realistic case and take $N_c=3, N_f=2$. By taking the parameter set Model IB with $\mu_G=0.43\rm{GeV}$  and $m_q=5\rm{MeV}, \sigma=(240\rm{MeV})^3$ as in our previous work \cite{Li:2013oda,Li:2012ay}, one can solve the equation of motions and obtain the numerical results. To investigate the effect
from the anomalous dimension, we choose three different cases: Case-I: Model IB in Ref.\cite{Li:2013oda,Li:2012ay}, with $\gamma=0$,  Case-II: Model IB with modelling $\gamma(z)$ taking the form of $\gamma(z)=-\frac{3}{2}(e^{-\Phi^2/2}-1)$, Case-III: Model IB with dynamical $\gamma(z)$ solving from running coupling, Case-IV: Model IB with a constant $\gamma=0.45$.

\subsection{Scalar and pseudoscalar mesons}
\begin{figure}[ht!]
    \centering
    \includegraphics[width=0.4\linewidth]{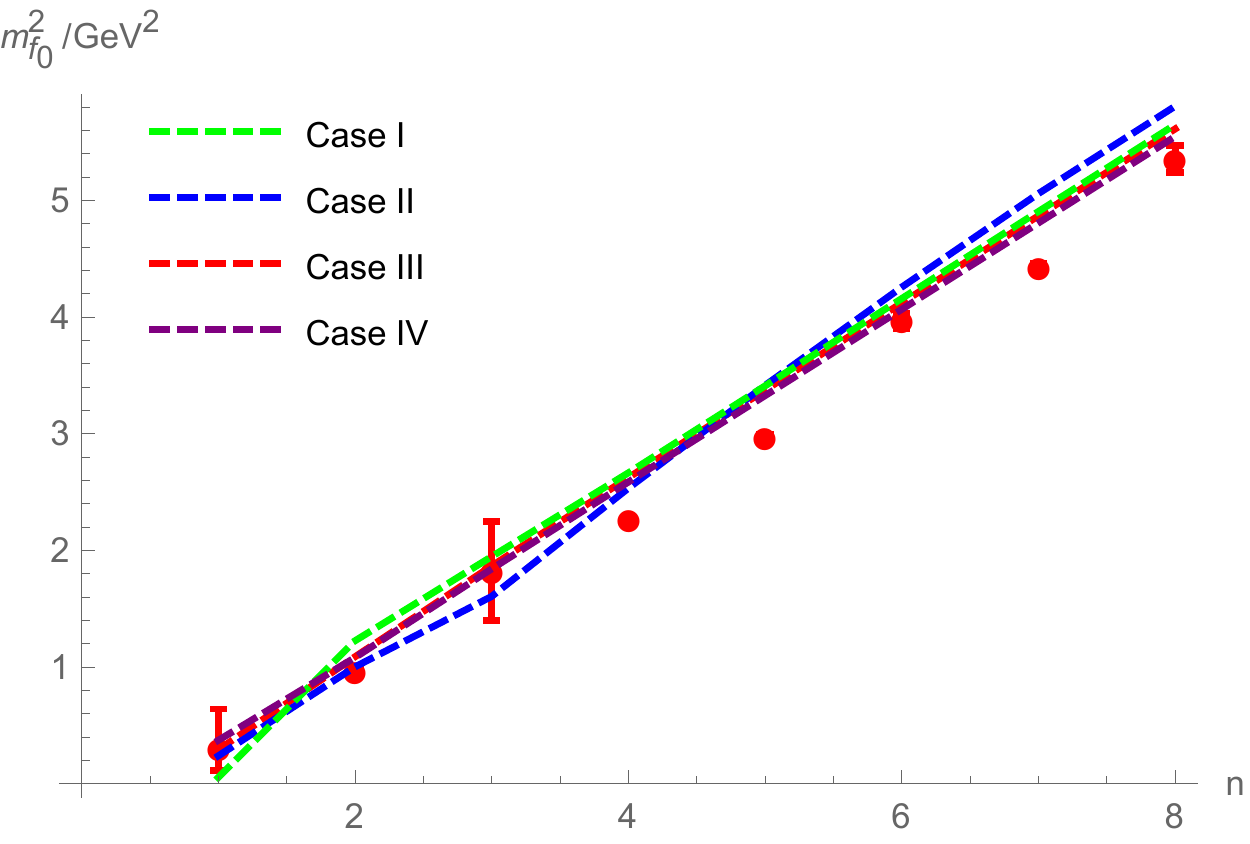}
    \hspace*{1cm}
     \includegraphics[width=0.4\linewidth]{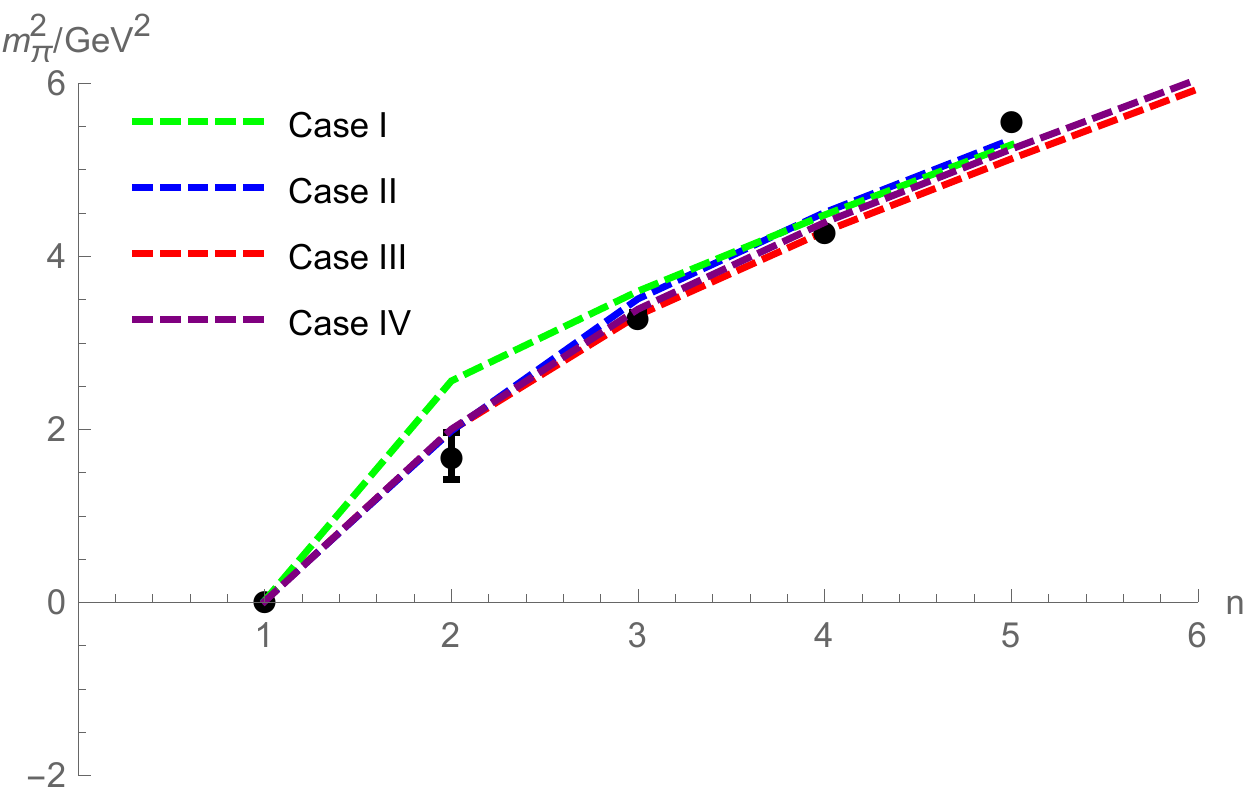}
     \vskip -0.05cm \hskip 0 cm
\textbf{( a ) } \hskip 8 cm \textbf{( b )}
    \caption{ Spectrum of scalars (a) and pseudoscalars (b). Case-I: Model-IB with $\gamma=0$, Case-II: Model-IB with model $\gamma(z)$ taking the form of $\gamma(z)=-\frac{3}{2}(e^{-\Phi^2/2}-1)$, and Case-III: Model-IB with dynamical $\gamma(z)$ solving from running coupling, Case-IV: Model IB with a constant $\gamma=0.45$. The red and black dots are experimental data, taken from Ref.~\cite{ParticleDataGroup:2008zun}, with a selection scenario following Ref.~\cite{Gherghetta:2009ac}.}
    \label{fig-scalar-sp}
\end{figure}

In this work, we consider the anomalous dimension of the $\bar{q}q$ operator only. Therefore, the most relevant sector would be the scalar sector. In this section we will study the corrections to the scalar and pseudo-scalar mesons, i.e. the $f_0$ meson and the pions. The numerical results for the ground states and excitation states for scalar meson $f_0$ and the pseudo-scalar meson pions $\pi$ are listed in Table \ref{scalarmasses}, \ref{pscalarmasses} and shown in Fig. \ref{fig-scalar-sp} for the above three cases. The parameter set Model IB (Case I) is fixed by fitting $m_{\pi}=140 {\rm MeV}$, and in the Model IB
the ground state mass of scalar is $231{\rm MeV}$, which is less than half of the experimental result. When the anomalous dimension for scalar field is taken into account,
the ground state mass of scalar is much improved as $488{\rm MeV}$, $539{\rm MeV}$, $609\rm{MeV}$ for Case-II, Case-III, Case IV respectively. It can also be read from Table \ref{scalarmasses}, \ref{pscalarmasses} and Fig. \ref{fig-scalar-sp} that the low excitation state masses in Model IB are much heavier than the experiment result, while in the case with the correction of the anomalous dimension for scalar field, the low excitation state masses are closer to the experiment results. It is interesting to see from Fig.\ref{fig-scalar-sp}(a) that, by a simple implement of the anomalous dimension, the scalar mass spectra could be much improved, especially for Case III when the anomalous dimension is solved dynamically.

For the pions, from Fig.\ref{fig-scalar-sp}(b), it is easy to see that similar improvement for the low lying states. The results with anomalous dimension is much closer to the experimental data for the low lying states. For the higher excitations, all the models tend to be give close results. But it should be pointed out that to show the effects of the anomalous dimension, we fix all the parameters as in Case I, in which the anomalous dimension is not added. Therefore, the results of the ground state of pions are different from the experimental data for Case II, III, and Case IV. The deviation for Case II is within $7\%$ and for Case III it is within $14\%$. For Case IV the deviation is larger, around $22\%$. For the second excitation states of pions, with the corrections from the anomalous dimension, the results of Case II, III, IV becomes much closer to the experimental data. For even higher excitations, the relative differences of all the cases become small. From the results we could see that though the anomalous dimension is not directly related with the pseudo scalar sector, it could also improve the model predicted spectrum, indirectly through modifying the background metric and field configuration.

\begin{table}
\begin{center}
\begin{tabular}{cccccccc}
\hline\hline
        n & $f_0$~Exp~(MeV)      & Case-I~(MeV) & Case-II~(MeV) &Case-III~(MeV) &Case-IV~(MeV) \\
\hline
        1 & $550^{+250}_{-150}$  & 231          &488            &539            &609   \\
        2 & $980 \pm 10$         & 1106         &1000           &1043           &1040    \\
        3 & $1350 \pm 150$       & 1395         &1267           &1366           &1357   \\
        4 & $1505 \pm 6$         & 1632         &1591           &1620           &1608    \\
        5 & $1873 \pm 7$         & 1846         &1848           &1837           &1824   \\
        6 & $1992 \pm 16$        & 2039         &2063           &2030           &2016    \\
        7 & $2103 \pm 8$         & 2215         &2249           &2206           &2192    \\
        8 &  $2314 \pm 25$       & 2376         &2410           &2367           &2355    \\
\hline\hline
\end{tabular}
\caption{The experimental and predicted mass spectra for scalar mesons $f_0$. Case-I: Model-IB with $\gamma=0$, Case-II: Model-IB with model $\gamma(z)$ taking the form of $\gamma(z)=-\frac{3}{2}(e^{-\Phi^2/2}-1)$, and Case-III: Model-IB with dynamical $\gamma(z)$ solving from running coupling, Case-IV: Model IB with a constant $\gamma=0.45$. The experimental data are taken from Ref.~\cite{ParticleDataGroup:2008zun}, with a selection scenario following Ref.~\cite{Gherghetta:2009ac}.} \label{scalarmasses}
\end{center}
\end{table}

\begin{table}
\begin{center}
\begin{tabular}{cccccccc}
\hline\hline
     &  n &$\pi$  Exp~(MeV)   & Case-I~(MeV) & Case-II~(MeV) &Case-III~(MeV) & Case-IV~(MeV) \\ \hline
      &  1 & $140$            & 140          &146            &158            &109\\
      &  2 & $1300 \pm 100$   & 1600         &1408           &1410           &1415\\
      &  3 & $1816 \pm 14 $   & 1897         &1873           &1821           &1841\\
      &  4 & $2070 $          & 2116         &2123           &2070           &2096\\
      &  5 & $2360 $          & 2299         &2313           &2264           &2289\\
\hline\hline
\end{tabular}
\caption{The experimental and predicted mass spectra for pseudoscalar mesons $\pi$. Case-I: Model-IB with $\gamma=0$, Case-II: Model-IB with model $\gamma(z)$ taking the form of $\gamma(z)=-\frac{3}{2}(e^{-\Phi^2/2}-1)$, and Case-III: Model-IB with dynamical $\gamma(z)$ solving from running coupling, Case-IV: Model IB with a constant $\gamma=0.45$. The experimental data are taken from Ref.~\cite{ParticleDataGroup:2008zun}, with a selection scenario following Ref.~\cite{Gherghetta:2009ac}.} \label{pscalarmasses}
\end{center}
\end{table}

\subsection{Vector and axial vector mesons}
\begin{figure}[ht!]
    \centering
    \includegraphics[width=0.4\linewidth]{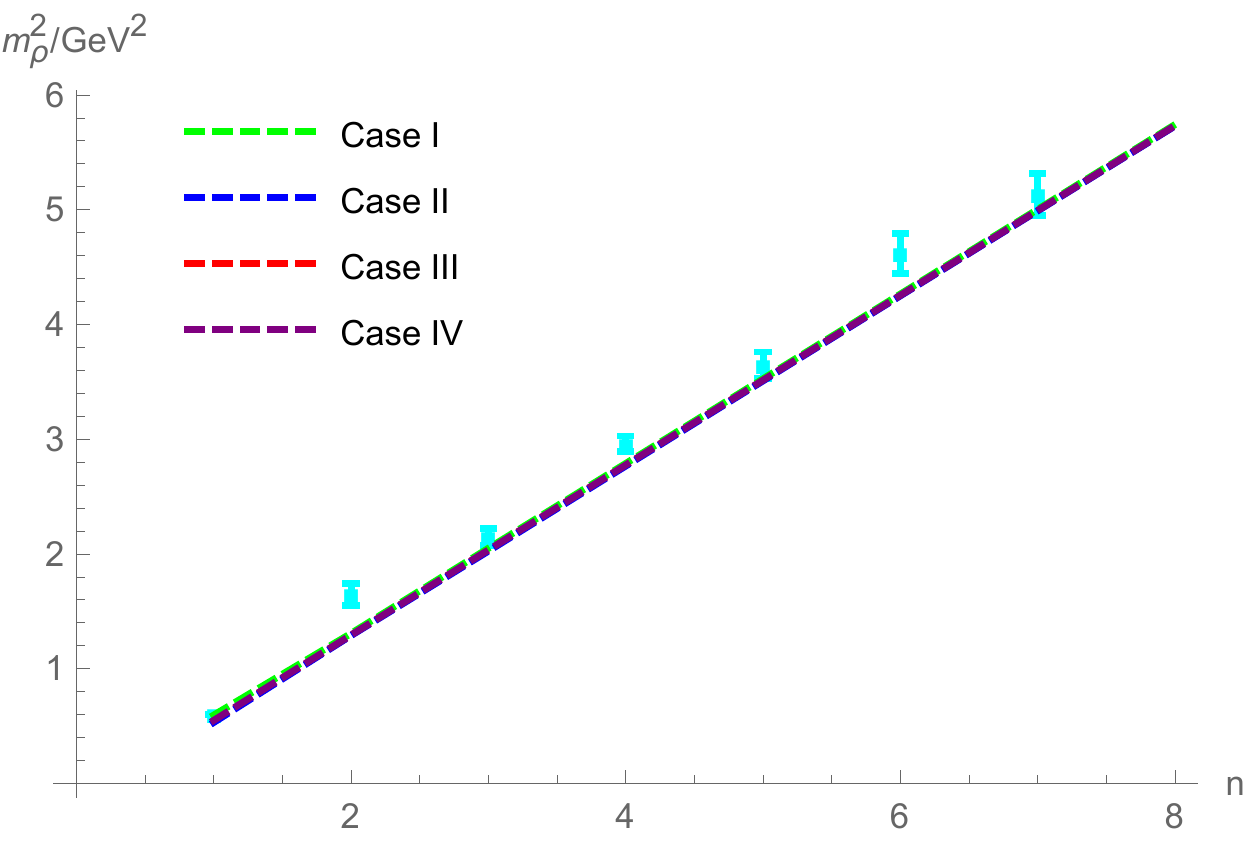}
    \hspace*{1cm}
    \includegraphics[width=0.4\linewidth]{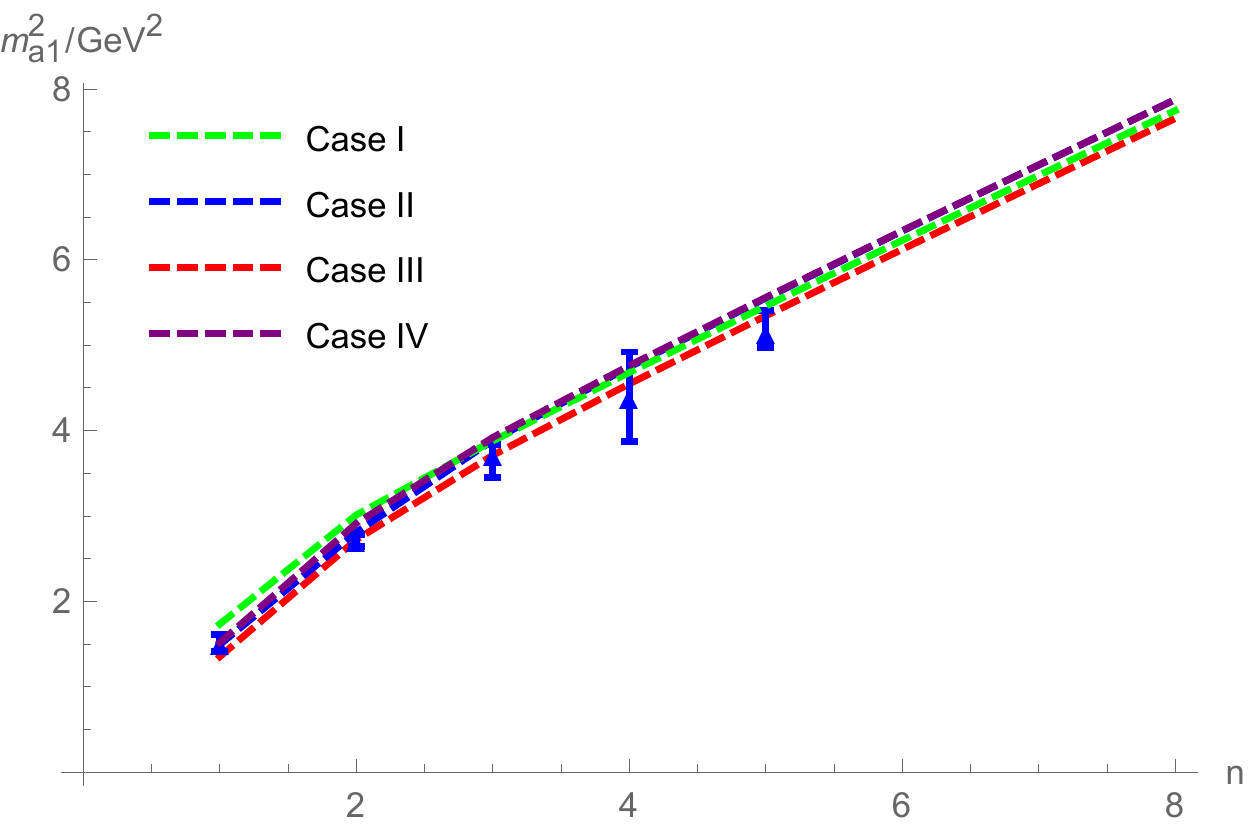}
      \vskip -0.05cm \hskip 0 cm
\textbf{( a ) } \hskip 8 cm \textbf{( b )}
    \caption{\label{fig-vectors} Spectrum of vector (a) and axial vector mesons (b). Case-I: Model-IB with $\gamma=0$, Case-II: Model-IB with model $\gamma(z)$ taking the form of $\gamma(z)=-\frac{3}{2}(e^{-\Phi^2/2}-1)$, and Case-III: Model-IB with dynamical $\gamma(z)$ solving from running coupling, Case-IV: Model IB with a constant $\gamma=0.45$. The cyan and blue dots are experimental data, taken from Ref.~\cite{ParticleDataGroup:2008zun}, with a selection scenario following Ref.~\cite{Gherghetta:2009ac}.}
    \label{fig-vector-sp}
\end{figure}

Then, we will consider the corrections from the anomalous dimension of the scalar operator to the vector and axial vector sectors, though this effect should be indirectly. The numerical results are given in Table \ref{vectormasses}, \ref{avectormasses} and Fig.\ref{fig-vectors}. From Table.\ref{vectormasses}, the anomalous dimension would reduce the masses of the lower excitations of $\rho$, but with only an amount of less than $7\%$. As for the higher excitations, the differences of different models are even smaller, less than $0.1\%$. This is mainly because the dominant effect for the $\rho$ spectra is the dilaton field, from which the linearity of the spectrum comes.

For the lower excitations of axial vector mesons, the corrections of the anomalous dimension decrease the masses, and the results becomes closer to the experimental data. From Fig. \ref{fig-vectors}(b), one could see that the results of Case III are much improved both in lower excitations and higher excitations.
\begin{table}
\begin{center}
\begin{tabular}{cccccccc}
\hline\hline
        n & $\rho$~Exp~(MeV) & Case-I~(MeV) & Case-II~(MeV) & Case-III~(MeV)  &Case-IV~(MeV) \\ \hline
        1 & $775.5 \pm 1$    & 771          &729            &737              &741\\
        2 & $1282 \pm 37$    & 1143         &1135           &1136             &1137\\
        3 & $1465 \pm 25$    & 1431         &1423           &1425             &1426\\
        4 & $1720 \pm 20$    & 1670         &1663           &1665             &1666\\
        5 & $1909 \pm 30$    & 1878         &1873           &1874             &1875\\
        6 & $2149 \pm 17$    & 2065         &2061           &2062             &2062\\
        7 & $2265 \pm 40$    & 2237         &2234           &2234             &2235\\
\hline\hline
\end{tabular}
\caption{The experimental and predicted mass spectra for vector mesons $\rho$. Case-I: Model-IB with $\gamma=0$, Case-II: Model-IB with model $\gamma(z)$ taking the form of $\gamma(z)=-\frac{3}{2}(e^{-\Phi^2/2}-1)$, and Case-III: Model-IB with dynamical $\gamma(z)$ solving from running coupling, Case-IV: Model IB with a constant $\gamma=0.45$. The experimental data are taken from Ref.~\cite{ParticleDataGroup:2008zun}, with a selection scenario following Ref.~\cite{Gherghetta:2009ac}.}
\label{vectormasses}
\end{center}
\end{table}

\begin{table}
\begin{center}
\begin{tabular}{cccccccc}
\hline\hline
        n & $a_1$~Exp ~(MeV)    & Case-I~(MeV)  & Case-II~(MeV) &Case-III~(MeV) &Case-IV~(MeV) \\
\hline
        1 & $1230 \pm 40$           & 1316      &1222           &1163           &1232\\
        2 & $1647 \pm 22$           & 1735      &1676           &1649           &1707\\
        3 & $1930^{+30}_{-70}$      & 1969      &1971           &1926           &1980\\
        4 & $2096 \pm 122$          & 2163      &2178           &2132           &2181\\
        5 & $2270^{+55}_{-40}$      & 2336      &2356           &2311           &2358\\
\hline\hline
\end{tabular}
\caption{The experimental and predicted mass spectra for axial
vector mesons $a_1$. Case-I: Model-IB with $\gamma=0$, Case-II: Model-IB with model $\gamma(z)$ taking the form of $\gamma(z)=-\frac{3}{2}(e^{-\Phi^2/2}-1)$, and Case-III: Model-IB with dynamical $\gamma(z)$ solving from running coupling, Case-IV: Model IB with a constant $\gamma=0.45$. The experimental data are taken from Ref.~\cite{ParticleDataGroup:2008zun}, with a selection scenario following Ref.~\cite{Gherghetta:2009ac}.} \label{avectormasses}
\end{center}
\end{table}

\subsection{Decay constants and form factors}

\begin{figure}[ht!]
    \centering
    \includegraphics[width=0.6\linewidth]{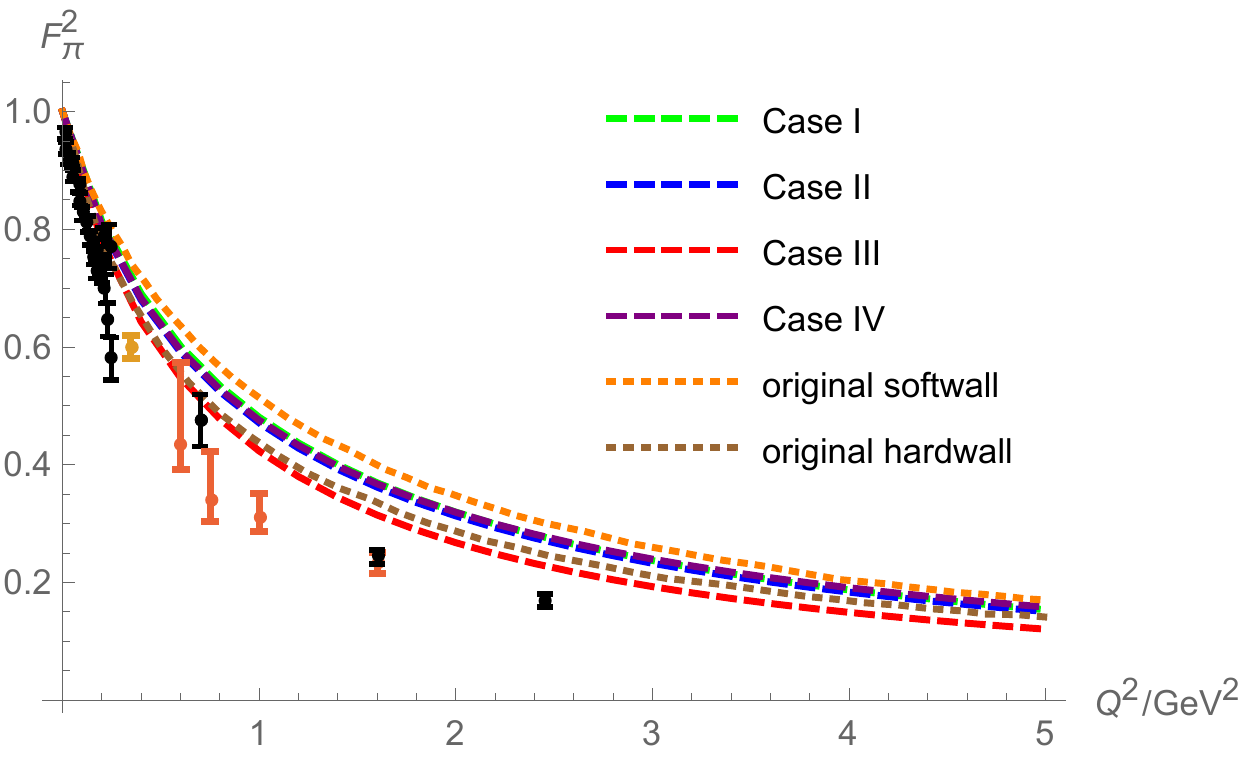}
    \caption{Formfactor $F_\pi$ as a function of $Q^2$ for different models. Case-I: Model-IB with $\gamma=0$, Case-II: Model-IB with model $\gamma(z)$ taking the form of $\gamma(z)=-\frac{3}{2}(e^{-\Phi^2/2}-1)$,and Case-III: Model-IB with dynamical $\gamma(z)$ solving from running coupling, Case-IV: Model IB with a constant $\gamma=0.45$.}
    \label{fig-formfactor}
\end{figure}

Using the formulae in Sec.\ref{formulae}, we can study the anomalous dimension corrections to the form factors and the decay constants. We list the results of decay constants and effective couplings in Table \ref{decay-coupling} and plot the results of form factors in Fig.\ref{fig-formfactor}. From Table \ref{decay-coupling}, we could see that with the anomalous dimension corrections, the decay constants of pion, $\rho$, $a_1$ becomes smaller. But the results of the effective coupling $g_{\rho\pi\pi}$ in Case III is closer to the experimental value.  The pion form factor $F_\pi(Q^2)$ has been investigated in light-front holographic QCD \cite{Li:2022mlg}, which was found to be in good agrreement with experimental data.

\begin{table}
\begin{center}
\begin{tabular}{cccccccc}
\hline\hline
 & ~Exp~(MeV)                     &Case-I~(MeV)  & Case-II~(MeV) &Case-III~(MeV)  &Case-IV~(MeV)  \\ \hline
 $f_\pi$        & $92.4\pm0.35$   & 83.6         & 81.0          &70.8            &80.3\\
 $F_\rho^{1/2}$ &$346.2\pm1.4$    &282           &267            &273             &273\\
 $F_{a_1}^{1/2}$&$433\pm13$       &452           &409            &406             &422\\
 $g_{\rho\pi\pi}$ &$6.03\pm0.07$  &3.14          &3.02           &3.77            &3.11\\
\hline\hline
\end{tabular}
\caption{Decay constants and couplings for different models. Case-I: Model-IB with $\gamma=0$, Case-II: Model-IB with model $\gamma(z)$ taking the form of $\gamma(z)=-\frac{3}{2}(e^{-\Phi^2/2}-1)$,and Case-III: Model-IB with dynamical $\gamma(z)$ solving from running coupling, Case-IV: Model IB with a constant $\gamma=0.45$. }
\label{decay-coupling}
\end{center}
\end{table}

\section{conclusion and discussion}
In this work, we introduce the anomalous dimension of the scalar operator $\bar{q}q$  in the dynamical holographic model, and study its corrections to the meson spectra, the decay constants, and the pion form factor. It is shown that with the anomalous dimension corrections, the background metric and the scalar field would recieve significant corrections, especially at the IR. The scalar meson spectrum would be much improved, especially for the low lying states. The vector meson spectra are almost not affected by the anomalous dimension of the scalar operator $\bar{q}q$. Only the ground state of the vector meson gets a correction within $7\%$. Since the anomalous dimension has significant effect on the background metric and fields, the masses of pions and the axial vector mesons also recieve important corrections, especially for the low lying states. In a short summary, the introduction of anomalous dimension could help improve the description of light mesons, especially for the scalar meson. In the current work, we only consider the anomalous dimension of scalar operator, it is also interesting to consider the corrections from other operators. But we will leave this part to the future.

\begin{acknowledgments}
 This work is supported in part by the National Natural Science Foundation of China (NSFC) Grant Nos. 12235016, 12221005, 11725523, 11735007 and 12275108, the Strategic Priority Research Program of Chinese Academy of Sciences under Grant Nos XDB34030000 and XDPB15, the start-up funding from University of Chinese Academy of Sciences(UCAS), and the Fundamental Research Funds for the Central Universities. Kazem Bitaghsir Fadafan is a PIFI visiting scientist in UCAS.
\end{acknowledgments}

\appendix

\section{The mass spectra and the Schrodinger-like equation}

Generally, the masses $m$ of hadrons could be extracted from the poles of the corresponding correlation functions, which take the form of  $G(p^2)\propto \frac{1}{p^2-m^2}$ with the momentum $p^2$ near the poles. Therefore, one can solve the correlation functions and get the masses from their poles. This is the standard way from the 4D field theory. As shown in \cite{Karch:2006pv}, in the holographic description, the solution of the poles can also be obtained by solving the egienvalues of the Schrodinger-like equations. In this appendix, we show the relation between the poles of the correlation functions and the egienvalues of the Schrodinger-like equation.

Here, we take the vector sector as an example. A basic assumption is that the 5D vector field $V_\mu^a$ is dual to the 4D operator $\bar{\psi}\gamma_\mu\tau^a\psi$. To get the correlation functions, one needs to obtain the partition function $Z[j_\mu(x)]$ with respect to the source $j_\mu^a$ (we will ignore the $SU(2)$ index $a$ in the following part, since it is irrelevant to the poles). From the dictionary, one can obtain the 4D correlation function from the 5D one and it has the form $Z[j(x)]=e^{iS_{onshell}[V(z=0,x)=j_\mu(x)]}$ applying the saddle point approximation. Then, the main task is to obtain the onshell action $S_{onshell}$ under the condition $V_\mu(z=0,x)=j_\mu(x)$.

With the metric ansatz Eq. \eqref{metric-ansatz}, one can easily derive the equation of motion for the vector field as
\begin{eqnarray}
\partial_z[e^{A_s-\Phi}\partial_z V_\mu(z,x)]+\partial_\nu\partial^\nu V_\mu(z,x)=0. \label{eom-vec}
\end{eqnarray}
and the onshell action as
\begin{eqnarray}\label{action-vec}
S=-\frac{1}{2g_5^2}\int dz d^4x\partial_z\big(
e^{3A_s-\Phi}V_\mu\partial_zV^{\mu}\big),
\end{eqnarray}
where we have insert the equation of motion into the action.
We can transform the 5D field $V_\mu$ to momentum space $V_\mu(z,x)\simeq\int d^4q \tilde{V}_\mu(z,q)e^{iq_\mu x^\mu}$. Then the expressions for the EOM and the on-shell action becomes:
\begin{eqnarray}
\partial_z[e^{A_s-\Phi}\partial_z \tilde{V}_\mu(z,q)]+q^2 e^{A_s-\Phi} \tilde{V}_\mu(z,q)=0,\\
S^{onshell}=-\frac{1}{2g_5^2}\int dz  d^4q\partial_z\big(
e^{3A_s-\Phi}\tilde{V}_\mu(z,-q)\partial_z\tilde{V}^{\mu}(z,q)\big).
\end{eqnarray}
According to the dictionary, we could take $\tilde{V}^{\mu}=V(z,q)j^{\mu}(q)$, with $j$ the 4D source depending on 4D momentum $q^\mu$ only and $V(z,q)$ representing the $z$ dependent part of $V^{\mu}$. Then, it is easy to see that $V(z,q)$ satisfies
\begin{eqnarray}
\partial_z[e^{A_s-\Phi}\partial_z V(z,q)]+q^2 e^{A_s-\Phi} V(z,q)=0.\label{eom-V1-1}
\end{eqnarray}
The onshell action then becomes
\begin{eqnarray}
S^{onshell}=-\frac{1}{2g_5^2}\int dz  d^4q\partial_z\big(
e^{A_s-\Phi}V\partial_z V\big)j_\mu j^\mu.
\end{eqnarray}
In order to make sure that the partition function depends on the 4D boundary quantities only, one has to impose additional boundary condition at $z=\infty$, i.e. $e^{A_s-\Phi}V\partial_zV|_{z=\infty}=0$. Following \cite{Karch:2006pv}, we would take $\partial_zV|_{z=\infty}=0$. So we can integrate over the z direction and get
\begin{eqnarray}
S^{onshell}=-\frac{1}{2g_5^2}\int  d^4q j_\mu j^\mu\big(
e^{A_s-\Phi}V\partial_z V\big)|_{z=\epsilon}.
\end{eqnarray}
In the asymptotic AdS region, $A(z)\simeq -\ln(z)$, and  we could get the near boundary condition of $V(z)$ as
\begin{eqnarray}
V(z,q)\rightarrow v_0(q)+v_2(q)z^2+... ~~~.
\end{eqnarray}
Here we have omitted the higher powers of $z$. Considering that at the boundary $z=0$, we have $V^\mu(z,q)=j^\mu(q)$, we have $V(z=0)=1$. We can normalize the solution with $v_0(q)$. Inserting it into the on-shell action, one gets
\begin{eqnarray}
S^{onshell}\simeq-\frac{1}{2g_5^2}\int  d^4q j_\mu^a(-q) j^{a,\mu}(q) \frac{v_2(q)}{v_0(q)}.
\end{eqnarray}
It is not difficult to see that the correlation function is proportional to
\begin{eqnarray}
G(q)\propto \frac{v_2(q)}{v_0(q)}.
\end{eqnarray}
Here, we shows the part relevant to the poles only. From this expression, the poles $q_n^2\equiv m_n^2$ satisfy $v_0(q=q_n)=0$.

Up to now, we have seen that, to get the poles, one does not need the full form of the correlation function. Instead, one can seak the solution of the masses $q_n^2$, which gives a solution satisfying Eq.\eqref{eom-V1-1} and the boundary conditions $V(z=0,q_n)=v_0=0$, $(V\partial_zV)|_{z=\infty}=0$.

Then, we can use another familiar form to get the poles $q_n^2$ or $m_n^2$.
In Eq.\eqref{eom-V1-1}, we can do a transformation
\begin{eqnarray}
V(z)=e^{\frac{A_s-\Phi}{2}}\psi(z).
\end{eqnarray}
Then Eq.\eqref{eom-V1-1} would becomes
\begin{eqnarray}
-\psi^{\prime\prime}(z)+V_{eff}(z)=q^2\psi(z),\\
V_{eff}(z)=\big(\frac{A_s^\prime-\Phi^\prime}{2}\big)^2+\frac{A_s^{\prime\prime}-\Phi^{\prime\prime}}{2}.
\end{eqnarray}
To get the poles, we should impose $V(z=0)=e^{A_s-\Phi}\psi(0)=\frac{\psi(z)}{z}|_{z=0}=0$, and $\partial_zV|_{z=\infty}=\partial_z(e^{-(A_s-\Phi)/2}\psi(z))|_{z=\infty}=0$. Actually, they are equivalent to the form used in many references $\psi(z=0)=\psi^\prime(z=\infty)=0$. Then, if we solve the above Schrodinger-like equation under those boundary conditions, we could get the same poles as solving Eq.\eqref{eom-V1-1}. For the other sectors, the derivation of the equivalence of the two method is quite similar, we would not repeat it here (for the scalar sector, one can refer to \cite{Cao:2021tcr,Cao:2020ryx}).

\bibliographystyle{unsrt}
\bibliography{refs}

\end{document}